\documentstyle[epsfig,twoside,fleqn,espcrc2]{article}

   \newcommand{\beq}{\begin{equation}} 
   \newcommand{\eeq}{\end{equation}}
\newcommand{\beqa}{\begin{eqnarray}} 
   \newcommand{\eeqa}{\end{eqnarray}}   
   \def\esim{\mathrel{\rlap{\raise2pt\hbox{$\sim$}}
    \lower1pt\hbox{$-$}}}         %equal to or approx. symbol
\newcommand{\bsg}{${\rm b}\to {\rm s}\gamma$}

\title{Non-Baryonic Dark Matter\thanks{
Invited talk at the 5th International Workshop on ``Topics in Astroparticle
and Underground Physics'' (TAUP97), Gran Sasso, Italy, 7-11 September, 1997,
to appear in the Proceedings, Nucl. Phys. Suppl., A. Bottino, A. di Credico and
P. Monacelli (eds.)}} 
\author{Lars Bergstr\"om\\
\vskip.2cm Department of Physics, 
Stockholm University \\ 
        P.O. Box 6730, SE-113 85 Stockholm, Sweden, lbe@physto.se}

\begin{document}

\begin{abstract}
The need for dark matter is briefly reviewed. A wealth of 
observational information points to the existence of a non-baryonic 
component. To the theoretically favoured 
candidates today belong axions, supersymmetric particles, and to some 
extent massive neutrinos. The theoretical foundation and experimental situation 
for each of these is reviewed. In particular, indirect detection 
methods of supersymmetric dark matter are described. Present experiments are just 
reaching the required sensitivity to discover or rule out some of these candidates, 
and major improvements are planned 
over the next few years. 
\end{abstract}

\maketitle

\section{Introduction}

The question of the nature of the dark matter in the Universe is 
getting more interesting than ever. As new observations are coming 
in, the possible candidates get more and more constrained. On the 
other hand, the picture which is emerging is to some extent 
puzzling, indicating that perhaps not all observations nor  
theoretical analyses are correct. 

Let us first recall that from the particle physics point of view, the
theoretically preferred (Einstein-De Sitter) 
Universe has the simple description 
\beq
\left\{
\begin{array}{ccc}
	\Omega & = & 1  \\
	\Lambda & = & 0\label{eq:eq1}
\end{array}\right.
\eeq
where $\Lambda$ is the cosmological constant, and
\beq
\Omega\equiv {\rho\over \rho_{crit}}={\rho\over 1.9\cdot 10^{-29}h^2\ 
{\rm g}\,{\rm cm^{-3}}},
\eeq
with $h$ related to the Hubble constant $H_{0}$ by $h=H_{0}/(100\ {\rm 
km}\,{\rm s}^{-1}{\rm Mpc}^{-1})$ (observationally, $h$ lies between 
0.4 and 0.8).

The cosmological model (\ref{eq:eq1}) has the attractive features that 
it is simple, avoids finetuning, and may be explained by a period of  
inflation in the earliest Universe.

Since Big Bang nucleosynthesis (BBN) puts an upper limit to
 the  baryonic contribution  $\Omega_{b}$ 
of  \cite{olive}
\beq
\Omega_{b}h^2\leq 0.026,\label{eq:bbn}
\eeq
non-baryonic dark matter dominates the energy density by a large 
factor in this type of model. 

Staying in the particle physicist's favourite Universe, one of the prime 
candidates for the non-baryonic component is provided by the lightest 
supersymmetric particle, plausibly the lightest neutralino $\chi$ (see below).

Supersymmetry seems to be a necessity in superstring theory (or 
M-theory) which unites all the 
fundamental forces of nature, including gravity. In most versions of 
the low-energy theory  there is a conserved multiplicative quantum 
number, R-parity, which makes the lightest supersymmetric particle 
stable. Thus, pair-produced neutralinos in the early Universe which 
left thermal equilibrium as the Universe expanded should have a non-zero 
relic abundance today. If the scale of supersymmetry breaking is related to 
that of electroweak breaking, $\Omega_{\chi}$ comes out in the right order 
of magnitude to explain the non-baryonic dark matter. Maybe it is 
asking too much of Nature, but it would indeed appear as an economic 
solution if  two of the  most outstanding problems in fundamental 
science, that of dark matter and that of the unification of the basic 
forces, would have a common element of solution - supersymmetry.

The BBN limit (\ref{eq:bbn}) is important, since it implies that if 
observations give a value of the total energy density above the BBN 
value, non-baryonic dark matter has to be present (or baryons have to 
be  hidden 
in some non-standard way at the time of nucleosynthesis), even if the 
total $\Omega$ turns out to be less than unity. Indeed, there are several 
independent indications that $\Omega > 0.1$ (and hardly any estimates 
at all that fall below that limit). Of course, it has long been 
recognized that even  the minimum value of $\Omega_{b}$ 
allowed by BBN is higher than the contribution from luminous baryons 
so that there also exists a dark matter problem for baryons - a lot of 
baryonic matter has to be hidden. Maybe the MACHO observations 
\cite{spiro} have a bearing on that problem.

Perhaps the strongest argument in favour of non-baryonic dark matter 
comes from structure formation and the microwave background. The 
basic picture is very simple: the observation of the isotropy of the 
microwave background to a level of a few times $10^{-5}$ 
through the COBE measurements, coupled with the theory of growth of 
perturbations in the theoretically simple linear regime, makes it essentially 
impossible to create the  non-linear structures we observe today with baryons 
only. In fact, the nice agreement between the COBE observations and 
the predictions from inflation of a nearly scale-invariant spectrum, 
may be taken as a piece of evidence in favour of inflation which 
could point to $\Omega=1$ on the largest scales. Recently, there has 
been a flurry of balloon and ground-based CMBR experiments on smaller 
angular scales, which probe the interesting dynamics of the acoustic 
peaks in the primordial cosmic fluid \cite{marc}. Although we have to 
await longer duration balloon flights and the MAP and Planck 
satellite missions for precision 
measurements, it seems that the present data  (interpreted with 
some courage) favour a  critical universe of $\Omega =1$ 
over an open Universe of, say, $\Omega=0.3$ \cite{lineweaver}.

Still on very large scales, analyses of the peculiar velocity ``flow'' 
of large clusters and other structures seem to need a lot of 
gravitating matter for its explanation, at least $\Omega > 0.4$ \cite{dekel}. 
The peculiar velocity field obeys the equation
\beq
\nabla\cdot {\bf v}={-\Omega^{0.6}\over 
b}\left({\rho-\langle\rho\rangle\over \langle\rho\rangle}\right),
\eeq
where $b=\delta\rho_{Gal}/\delta\rho_{M}$ is the ``biasing'' parameter 
which tells how light traces mass. The combination $\Omega^{0.6}/b$ is 
determined by the analysis of \cite{dekel} to be $0.89\pm 0.12$, which 
is  consistent with $\Omega=1$, $b=1$. Using the theoretical limit
$b>0.75$, a 95~\% c.l. limit of $\Omega>0.33$ can be given.

On scales up to a redshift around unity, gravitational lensing 
\cite{kochanek} and deep supernova searches
\cite{perlmutter,kirshner} provide interesting new methods which, however, still need 
to be improved as regards the systematic errors. 
Indeed, the first 7 supernovas analysed in \cite{perlmutter} imply a 
large value for the matter density $\Omega_{M}$ (and a small value of 
the vacuum energy contribution $\Omega_{\Lambda}$),
 whereas the 
4 supernovas of \cite{kirshner} favour a rather smaller $\Omega_{M}\sim 0.3 $ (and is not 
incompatible with $\Omega_{M}+\Omega_{\Lambda}=1$). This is a field 
of great potential which evolves rapidly, and the use of the 
infrared camera on the Hubble Space Telescope should make follow-up 
observations of high-$z$ objects easier. 

The gravitational lensing analysis of \cite{kochanek} indicates that 
there is plenty of dark matter; the 95~\% c.l. limits are 
$\Omega_{M}>0.38$ and $\Omega_{\Lambda}<0.66$. An analysis of the 
number of arcs from gravitational lensing of clusters expected in 
various cosmologies gives consistency for an open model with 
$\Omega\sim 0.3-0.4$, but failure for closed models with or without 
a cosmological constant \cite{bartelmann}.

The analysis of galaxy clusters  has not yet converged to a universal 
value of $\Omega$. There are some indications \cite{blanchard} from 
the temperature-luminosity relation for rich clusters that a high 
value ($\Omega\sim 1$) might be needed. On the other hand 
\cite{sasha,carlberg} other dynamical estimates are more consistent 
with a lower value, $\Omega\sim 0.2-0.3$. 

On galactic scales and smaller, the classical tests of the mass 
distribution provided by rotation curves continue to be refined. A 
recent compilation of almost 1000 rotation curves led to the 
conclusion that dark matter indeed is present in large amounts \cite{persic}. 
A very interesting class of 
objects is provided by low surface brightness galaxies and
the dwarf spheroidal satellite galaxies to the 
Milky Way, which seem to be completely dominated by dark matter 
\cite{gilmore}. A couple of these have unusual rotation curves which 
could perhaps be interpreted as being due to a combination of MACHOs 
and nonbaryonic dark matter \cite{burkert_silk}. 

The problem of how 
dark matter is distributed in halos of galaxies and galaxy clusters 
is an important one for the purpose of determining strategies for the 
detection of the various candidates, as we will see. Unfortunately, 
the available data on the structure of the Milky Way do not constrain 
the dark matter halo density profile very much \cite{dehmelt}.

A problem for high-$\Omega$ models without cosmological constant was 
until very recently the difficulty of reconciling the age of the 
Universe $t_{U}$ based on the present expansion rate $H_{0}$
with the estimated age of the oldest globular clusters. The values 
$\Omega_{M}=1$, $\Omega_{\Lambda}=0$  give
$t_{U}=2/(3H_{0})$, which for $h=0.6$ implies $t_{U}\sim 10-11$ Gyr. The 
determination of globular cluster ages on the other hand  used to give  
$14-15$ Gyr as best estimates. Besides some doubts that may still remain 
about the accuracy of these latter very indirect means of bounding
the age of the Universe, it seems that the recalibration of the 
distance scale provided by the recent Hipparcos satellite  parallax 
measurements brings the globular cluster age limit down by 2-3 Gyr 
\cite{chaboyer}, with the one-sided 95 \% c.l. lower limit being 9.5 
Gyr. 
This means that  a critical universe is now allowed without 
cosmological constant, if $h\leq 0.67$, a value that is not far from 
the current best estimates.

To summarize at this point: A variety of independent estimates of the 
matter density in the Universe point to a value larger than the 
maximal value provided by baryons alone according to nucleosynthesis. 
The need for nonbaryonic dark matter is therefore striking. If the 
``natural'' theoretical prediction $\Omega_{M}=1$, 
$\Omega_{\Lambda}=0$ is fulfilled is a different question, for which 
most of the observational data today do not yet give support, except 
maybe at the largest scales.

\section{Dark Matter Candidates}

Given that the total mass density of the universe seems to be higher 
than what is allowed by Big Bang nucleosynthesis, an important task of 
cosmology and particle physics is to produce viable non-baryonic 
candidates and to indicate how the various scenarios can be tested 
observationally. 

\subsection{Baryons}

The ``second'' dark matter problem, to account for the baryons that 
have to hidden in order to get agreement with BBN, has not been fully 
solved. It is possible that a lot of the baryonic mass may be hidden 
in galactic halos in the form of sub-solar mass objects, MACHOS 
\cite{spiro}. However, even with the surprisingly large optical depth 
for microlensing observed towards the LMC, the most likely fraction of 
the halo mass given by MACHOs is not larger than 50 \% and could in 
fact be much smaller if debris from tidal stripping of the LMC itself 
or other dwarf satellites happens to lie in the line-of-sight, as 
indicated by some observations \cite{zaritsky}.

Probably, the main repository of baryons in the universe is the gas 
of rich clusters. These systems are large enough that the baryon 
fraction should be a good tracer of the total $r_{b}=\Omega_{b}/\Omega_{M}$. 
Estimates \cite{clusterfraction} give a value of $r_{b}\sim 0.1 - 
0.2$, 
which combined with the BBN determination of $\Omega_{b}$ gives 
$\Omega_{M}\sim 0.1$ if the high deuterium measurement \cite{highd}
is correct, 
$\Omega_{M}\sim 0.5$ for the low deuterium abundance case 
\cite{lowd}, with probably rather large systematic uncertainties 
related to the limited understanding of how clusters formed.

\subsection{Neutrinos}

Of the many candidates for non-baryonic dark matter proposed, 
neutrinos are often said to  have the undisputed virtue of being known to exist. 
Actually, this is  a statement which needs some qualification because 
neutrinos can only be dark matter candidates if they are massive. For 
this to be true, both left-handed and right-handed neutrino states 
are needed, and the latter are not known to exist (in the minimal 
Standard Model of particle physics the right-handed neutrino is simply 
absent). In principle, one can construct a mass term from only the 
left chirality neutrino field, but this give a Majorana type mass 
which violates lepton number by two units, and in the Standard Model
$B-L$ is exactly conserved. 

Non-zero neutrino masses, if established, would thus be an indication 
of physics beyond the Standard Model. Since there exists a number of 
indications that the Standard Model cannot be the final theory, it 
would not be a big surprise if neutrinos are massive. As the direct 
experimental limits on neutrino mass show \cite{pdg}:
\begin{center}
\beqa
m_{\nu_{e}}&<&15\ {\rm eV}\nonumber\\
m_{\nu_{\mu}}&<&0.17\ {\rm MeV}\nonumber\\
m_{\nu_{\tau}}&<&24\ {\rm MeV}\nonumber\\
\eeqa
\end{center}
the neutrino masses have to be much smaller than the corresponding 
quark and charged-lepton masses. An intriguing explanation of this 
fact could be given by the so-called see-saw mechanism, 
where a right-handed Majorana mass $M$, at a large scale $\propto 
M_{GUT}\sim 10^{15\pm 2}$ GeV  modifies through mixing the usual Dirac-type mass 
$m_{D}$ of the lightest state  to $m_{D}^2/M\ll m_{D}$. In the 
simplest versions of this 
scheme, the neutrino masses would scale as the square of the 
corresponding charged-lepton masses. There are variants (e.g. in 
models of loop-induced neutrino masses) where neutrino masses 
are instead linearly related to the charged-lepton masses.

Indeed, there are several indications that neutrinos are not 
massless. Although none of the direct kinematical measurements of 
neutrino masses has given a value inconsistent with zero,  evidence 
from neutrino oscillation experiments is mounting that neutrinos 
oscillate in flavour and hence must posses non-zero masses. To give a 
cosmologically interesting contribution to $\Omega$, a relatively 
narrow range $m_{\nu}\sim 1-50$ eV is required. A neutrino heavier 
than that would overclose the universe (unless $m_{\nu}> 3$ GeV, 
which for Dirac neutrinos is ruled out by accelerator and direct detection data
up to the TeV range), whereas 
a lighter neutrino would only give a small and dynamically not very
important contribution to $\Omega$. 

Of the various experimental indications of neutrino oscillations, 
only the LSND results \cite{caldwell} seem to be in the cosmologically 
interesting range, with $\Delta m^2\sim 1 - 6$ eV$^2$. These results, however, 
need independent 
confirmation from other experiments. 

The solar neutrino problem, 
which in view of new helioseismological data 
does not seem to be solvable by tampering with the solar model
\cite{dalsgaard}, 
and thus presents rather compelling evidence for oscillations, only 
gives solutions with very small $\Delta m^2$. The weak indications of 
an energy-dependence of the solar neutrino deficit seen so far in the 
Super-Kamiokande data \cite{totsuka} favour the small-angle MSW 
solution, which has $\sin^22\theta\sim 5\cdot 10^{-3}$ 
and $\Delta m^2\sim 10^{-5}$ eV$^2$.
 This would indicate 
small absolute values of neutrino masses unless there would be mass
degeneracies of unknown origin between neutrinos. 

Likewise, the 
atmospheric neutrino anomaly, recently confirmed by Super-Kamiokande 
data, has a preferred solution with a $\Delta m^2$ of only a few 
times $10^{-3}$ eV$^2$. Seen already in the smaller Kamiokande 
detector, as well as in IMB and Soudan-2 \cite{goodman} as a deficit 
in the ``ratio of ratios'', $r\equiv (\nu_{\mu}/\nu_{e})_{data}
/ (\nu_{\mu}/\nu_{e})_{MC} \sim 0.6$, the interpretation in terms of 
neutrino oscillations seems much stronger with the Super-Kamiokande 
data where a zenith-angle dependence of the ratio is indicated 
\cite{totsuka} with higher significance than in Kamiokande. 
Thus it seems plausible that neutrinos are 
indeed massive, but the mass is too small to be very significant for 
cosmology. 

One should keep in mind, however, the need to check the LSND 
claims  of a neutrino mass difference in the eV region, which if 
found correct could give a $10-20$ \% contribution to 
$\Omega_{M}$. Such 
a mixture of hot and cold dark matter makes the galaxy and cluster
large-scale structure power spectrum 
easier to connect to the COBE measurements of the Cosmic Microwave 
Background Radiation (CMBR). Also, even if the largest neutrino mass 
is of the order of only a few tenths of an eV, as indicated by the 
atmospheric neutrino anomaly, the effects on structure formation could 
still be large enough to be detected, e.g., by the Sloan Digital Sky 
Survey \cite{huetal}.

There is an additional  fundamental objection to having neutrinos as 
the dominant 
constituent of
dark matter on all scales where it is observationally needed. This 
has to do with the fact that neutrinos are spin-$1/2$ particles 
obeying the Pauli exclusion principle. To make up the dark matter 
in dwarf galaxies (which in fact are seen to be completely dominated 
by the dark matter component), neutrinos would have to be stacked 
together so tightly in phase-space that it seems difficult to evade 
the Pauli principle. Quantitatively, Tremaine and Gunn found \cite{tg}
that to explain the dark matter of a dwarf galaxy of velocity 
dispersion $\sigma$ and core radius $r_{c}$, the neutrino mass has to 
fulfil
$$
m_{\nu}\geq 120\ {\rm eV}\left({100\ {\rm km/s}\over\sigma}\right)^{{1\over 
4}}\left({1\ {\rm kpc}\over r_{c}}\right).
$$

This high value is, however, not consistent with the requirement 
$\Omega_{\nu}h^2\leq 1$, which requires $\sum_{i}m_{\nu_{i}}< 90$ eV.

\subsection{Axions}

Axions are hypothetical particles, spinless light pseudoscalar 
bosons, which appear in models which explain the smallness of the 
CP violating $\theta$ parameter of QCD by the existence of a global 
symmetry, $U(1)_{PQ}$, which is spontaneously broken. The Goldstone 
boson of this symmetry breakdown is the axion, which however gets a 
non-zero mass from the QCD anomaly (which can be interpreted as a 
mixing of the axion field with the $\pi$ and $\eta$ mesons). The 
phenomenology of the axion is determined, up to numerical factors, by 
one number only - the scale $f_{a}$ of symmetry breaking. In 
particular, the mass is given by
\beq
m_{a}=0.62\ {\rm eV}\left({10^7\ {\rm }\over f_{a}}\right),
\eeq
and the experimentally important coupling to two photons is due to 
the effective Lagrangian term
\beq
{\cal L}_{a\gamma\gamma}=\left({\alpha\over 2\pi f_{a}}\right)\kappa{\bf E\cdot B}a,
\eeq
where $\kappa$ is a model-dependent parameter of order unity. 

The axion, constrained by laboratory searches, stellar cooling 
 and the dynamics of supernova 1987A to be very light, $m_{a}<$ (few 
 eV) \cite{raffelt}, couples so weakly to other matter that it would 
 behave today as Cold Dark Matter. The window where axions are viable 
 DM candidates is progressively getting smaller, but still there is an 
 acceptable range between around $10^{-5}$ and $10^{-2}$ eV where they 
 pass all observational constraints and would not overclose the Universe.
 Fortunately, there are now  two experiments \cite{llnl,kyoto} 
 which have the experimental sensitivity of probing much of the remaining 
 window within the next few years.
 
 \subsection{Other Solutions to the Dark Matter Problem}
 
 Axions share with massive neutrinos and the supersymmetric candidates to be 
 discussed next the attractive feature of having other, 
 particle-physics motivated, reasons to exist besides giving a 
 possible explanation of dark matter. Of course there are other 
 proposed candidates which, although not yet generally accepted, could 
 finally turn out to give the correct explanation. 
 
 It has 
 turned out to be very difficult to modify gravity on the various length 
 scales where the dark matter problem resides, but one cannot logically 
 exclude the possibility that this could be finally achieved. 
 
 The 
 three classes of non-baryonic dark matter we discuss here have the 
 additional virtue of lending themselves to experimental 
 investigations at a level that is already starting to probe relevant 
 regions of parameter space.
 
 The next class of models, supersymmetric (susy) dark matter, should be seen 
 at one particular realization of a generic type of models, sometimes 
 named WIMPs (weakly interacting massive particles). Here weakly 
 interacting, electrically neutral massive (GeV to TeV range) 
 particles are assumed to carry a conserved quantum number ($R$-parity 
 in the case of susy) which 
 suppresses or forbids the decay into lighter particles. Such 
 particles should have been copiously produced in the early universe 
 through their 
 weak interactions with other forms of matter and radiation. As the 
 universe expanded and cooled, the number density of the WIMPs 
 successively became too low for the annihilation processes to keep 
 up with the Hubble expansion rate. A relic population of WIMPs should 
 thus exist, and it is very suggestive that the canonical weak 
 interaction strength is, according to detailed calculations, 
 just right to make the relic density fall in the 
 required range to contribute substantially to $\Omega$.
 
 In addition, WIMPs are generically found to decouple at a temperature 
 that is roughly $m_{\rm WIMP}/20$, which means that they are 
 non-relativistic already at decoupling and certainly behave as CDM 
 by the time of matter dominance and structure formation. 
 
 Although the details of structure formation probably will remain unclear 
 until the next generation of microwave background measurements and 
 digital sky surveys are available, the formerly so popular ``Standard CDM'' 
 (SCDM) model with $\Omega_{CDM}=0.95$, $\Omega_{b}=0.05$, 
 the slope parameter of the scale invariant primordial power 
 spectrum $n=1$ , $h=0.5$  seems to be more or less ruled out by observations.
 The main problem is that normalization to the COBE spectrum at the 
 largest scales causes by a factor of 2 or so too much power on the 
 smaller scales probed by galaxy and cluster surveys. 
 However, with only small modifications such as adding an HDM 
 component, tilting the primordial spectrum to $0.8-0.9$, decreasing 
 $h$ or a combination thereof one can get a very satisfactory 
 description of essentially all the data. As these are all very 
 reasonable modifications (and indeed SCDM an oversimplified model), 
 the case for a large component of the CDM type appears as strong as 
 ever.  
\subsection{Supersymmetric particles}

One of the  favoured particle dark matter candidates is the lightest
supersymmetric particle $\chi$, assumed to be a neutralino, i.e. a mixture 
of the supersymmetric partners of the photon, the $Z$ and the two neutral
$CP$-even Higgs bosons present in the minimal extension of the 
supersymmetric standard model (see, e.g. \cite{haberkane}). The 
attractiveness
of this candidate stems from the fact that its generic couplings
 and mass range
naturally gives a relic density in the required range to explain halo 
dark matter. Besides, its 
motivation from particle physics has recently become stronger due to 
the apparent need for 100 GeV - 10 TeV scale supersymmetry to achieve
unification of the gauge couplings in view of LEP results
\cite{amaldi}. (For a recent review of supersymmetric dark 
matter, see Ref.\,~\cite{jkg}.) 

Thanks to new exciting developments in string theory \cite{strings}, 
supersymmetry has become an even more attractive feature to be 
expected at the doorstep beyond the Standard Model. At a more 
phenomenological level, supersymmetry gives a nice solution to the 
so-called hierarchy problem, which is to understand why the 
electroweak scale is so much smaller than the Planck scale despite the 
fact that there is nothing in non-supersymmetric theories to 
cancel the severe quadratic divergences of loop-induced mass terms. 
In supersymmetric theories, the partners of differing spin would 
exactly cancel those divergencies (if supersymmetry were unbroken).

\subsubsection{MSSM: The minimal supersymmetric extension of the standard model}

The minimal $N=1$ supersymmetric extension
of the standard model is defined by the the particle content and
gauge couplings required by supersymmetry and a gauge-invariant
 superpotential. The only addition to the obvious doubling of the 
 particle spectrum of the Standard Model concerns the Higgs sector. It 
 turns out that the single scalar Higgs doublet is not enough to give 
 masses to both the $u$- and $d$-like fermions and their 
 superpartners. Thus, two complex Higgs doublets have to be 
 introduced. After the usual Higgs mechanism, three of these states 
 disappear as the longitudinal components of the weak gauge bosons 
 leaving 5 physical states: two neutral scalar Higgs particles $H_{1}$ and 
 $H_{2}$ (where by convention $H_{2}$ is the lighter state), one 
 neutral pseudoscalar state $A$, and two charged scalars $H^{\pm}$.
 The $Z$ boson mass gets a contribution from the vacuum expectation 
 values (VEVs) of both of the doublets, but the way this division is done 
 between the VEV $v_{1}$ of $H_{1}$ and $v_{2}$ of $H_{2}$ is not fixed a priori.
 
 Electroweak symmetry breaking is thus caused by both $H_1$ and $H_2$
acquiring vacuum expectation values,
\begin{equation}
  \langle H^1_1\rangle = v_1 , \qquad \langle H^2_2\rangle = v_2,
\end{equation}
with $g^2(v_1^2+v_2^2) = 2 m_W^2$, with the further assumption that
vacuum expectation values of all other scalar fields (in particular,
squark and sleptons) vanish. This avoids color and/or charge breaking
vacua. 
 The ratio of VEVs
 \beq
 \tan\beta\equiv {v_{2}\over v_{1}}
 \eeq
 always enters as a free parameter in the MSSM, although it seems 
 unlikely to be outside the range between 1.1 and 45 \cite{jkg}.
 
 After supersymmetrization, the theory  also has to contain the 
 supersymmetric partners of these spin-0 Higgs fields. In particular, 
 two Majorana fermion states, higgsinos,  are the supersymmetric 
 partners of $H_{1}$ and $H_{2}$. These can mix with each other and 
 with two other neutral
 Majorana states, the supersymmetric partners of the photon (the photino)
 and the $Z$ (the zino). When diagonalizing the mass matrix of these 
 four neutral Majorana spinor fields (neutralinos), the lightest physical state
 becomes an excellent candidate for Cold Dark Matter.
 
 The non-minimal character of the Higgs sector may well be the first 
 experimental hint at accelerators of supersymmetry. At tree level, 
 the $H_{2}$ mass is smaller than $m_{Z}$, and even after allowing for 
 radiative corrections it can hardly be larger than around 140 GeV.
 
\subsubsection{Supersymmetry Breaking}

Supersymmetry is a mathematically beautiful theory, and would give 
rise to a very predictive scenario, if it were not broken in an 
unknown way which unfortunately introduces a large number of unknown 
parameters.

Breaking of supersymmetry has of course to be present 
since no supersymmetric particle has as yet been 
detected, and supersymmetry requires
particles and sparticles to have the same mass. This breaking can be
 achieved in the MSSM by a soft
supersymmetry-breaking potential which does not re-introduce large
 radiative mass-shifts (and which strongly indicates that the 
 lightest supersymmetric particles should not be too much heavier than the 250 
 GeV electroweak breaking scale). The origin of this effective 
 low-energy  potential need not be specified, 
 but it is natural to believe that it is induced through explicit 
 breaking in a hidden sector of the theory at a high mass scale. The 
 susy breaking terms are then transmitted to the visible sector 
 through gravitational interactions.
 
 Another possibility, of recent resurging interest, is that 
 supersymmetry breaking is achieved through gauge interactions at 
 relatively low energy in the hidden sector \cite{gauge}. This is then 
 transferred to the visible sector through some messenger fields which 
 transform non-trivially under the Standard Model gauge group. 
 Although this scenario has some nice features, it does not  seem to give 
 as natural a candidate for the dark matter as the ``canonical'' 
 scenario, which is the one we shall assume in most of the following.
 See, however, \cite{gaugedm} for some possibilities of dark matter 
 candidates in gauge-mediated models.
  
 Since one of the virtues of supersymmetry is that it resurrects the 
 hope for a grand unification at a common mass scale,
  a simplifying unification assumption is often used
for the gaugino mass parameters,
\begin{equation}
  \begin{array}{l}
  M_1 = {5\over 3}\tan^2\theta_wM_2\esim 0.5 M_2, \\
  M_2 = { \alpha_{ew} \over \sin^2\theta_w \alpha_s } M_3 \esim 0.3 M_3.
  \end{array}
  \label{gauginounif}
\end{equation}
As mentioned, the one-loop effective potential for the Higgs fields 
 has to be used used to obtain  realistic mass estimates.
 The minimization conditions of the potential allow one to
trade two of the Higgs potential parameters 
for the $Z$ boson mass $m_Z^2 = {1\over2} (g^2+g'^2)
(v_1^2+v_2^2)$ and the ratio of VEVs $\tan\beta$. The third parameter 
can further be reexpressed in terms of the mass of one of the physical
Higgs bosons, for example $m_{A}$.

The neutralinos $ \tilde{\chi}^0_i$ are linear combination of the
neutral gauge bosons ${\tilde B}$, ${\tilde W_3}$ (or equivalently
$\tilde\gamma$, $\tilde Z$) and of the neutral
higgsinos ${\tilde H_1^0}$, ${\tilde H_2^0}$.  In this basis, their
mass matrix
\begin{eqnarray}
  {\cal M} = 
  \left( \matrix{
  {M_1} & 0 & -{g'v_1\over\sqrt{2}} & +{g'v_2\over\sqrt{2}} \cr
  0 & {M_2} & +{gv_1\over\sqrt{2}} & -{gv_2\over\sqrt{2}} \cr
  -{g'v_1\over\sqrt{2}} & +{gv_1\over\sqrt{2}} & 0 & -\mu \cr
  +{g'v_2\over\sqrt{2}} & -{gv_2\over\sqrt{2}} & -\mu & 0 \cr
  } \right)
\end{eqnarray}
can be diagonalized  to give four neutral Majorana states,
\begin{equation}
  \tilde{\chi}^0_i = 
  a_{i1} \tilde{B} + a_{i2} \tilde{W}^3 + 
  a_{i3} \tilde{H}^0_1 + a_{i4} \tilde{H}^0_2 
\end{equation}
the lightest of which,  $\chi$, is then the candidate for
the particle making up (at least some of) the dark matter in the universe.

For simplicity, one often makes a  
diagonal ansatz for the 
 soft supersymmetry-breaking parameters in the sfermion sector.
This allows the squark mass matrices to be diagonalized analytically.
Such an ansatz implies the absence of
tree-level flavor changing neutral currents (FCNC) in all sectors of the
model. In models inspired by low-energy supergravity with
a universal scalar mass at the grand-unification (or Planck) scale
 the running of the scalar masses down to the electroweak scale
generates off-diagonal terms and tree-level FCNC's in the squark
sector. For a discussion of this class of models, and of effects 
related to relaxing the assumption of universal scalar masses, see 
\cite{ellis}. In most of the estimates of detection rates given 
below, we will adhere to a purely phenomenological approach, where the 
simplest unification and scalar sector constraints are assumed, but no 
supergravity relations are used.

When using the minimal supersymmetric standard model 
in calculations
 of relic dark matter density, one should make sure that all
accelerator  constraints on supersymmetric particles and couplings are
imposed. In addition to significant restrictions on parameters given by
LEP \cite{lepbounds}, the measurement of the \bsg\ process 
is providing important bounds.

The relic density calculation in the MSSM for a given set of 
parameters is nowadays accurate to 10~\% or so. A recent 
important improvement is the inclusion of 
coannihilations, which can change the relic abundance by a large 
factor in some instances \cite{coann}.

\subsubsection{Detection methods}

If neutralinos are indeed the CDM needed on galaxy scales and larger, 
there should be a substantial flux of these particles in the Milky 
Way halo. Since the interaction strength  is 
essentially given by the same weak couplings as, e.g., for neutrinos 
there is a non-negligible chance of detecting them in low-background 
counting experiments. Due to the large parameter space of MSSM, even 
with the simplifying assumptions above, there is a rather wide span of 
predictions for the event rate in detectors of various types. It is 
interesting, however, that the models giving the largest rates are 
already starting to be ruled out by present direct detection 
experiments \cite{bsg,bottino}.

Besides these possibilities of direct detection of supersymmetric dark
 matter, discussed extensively at this Workshop \cite{caldwell}
 (indeed even with a 
 weak indication of a positive signal \cite{dama,fiorenza}), 
 one also has the possibility of indirect detection through 
 neutralino annihilation 
in the galactic halo. This is becoming a promising method thanks
 to very powerful new detectors for 
cosmic gamma rays  and neutrinos planned and under construction.
 
%In principle, all stable particles produced in annihilation processes in 
%the halo can serve 
%as signatures of the dark matter candidate. However, 
%electrons and protons are much 
%too abundant in the ordinary cosmic rays to be useful. 
%Much lower background fluxes, 
%and therefore greater potential for detection of an additional 
%component, are present in the case of positrons and antiprotons. 
%The problem is still that the signal does not have 
%a distinct signature. This is due to the fact that the primary 
%annihilation processes are into quarks,  heavy leptons, gauge bosons and 
%Higgs particles, whereas the positrons 
%and antiprotons are secondary or tertiary decay products. 
%The reason for the small 
%direct coupling to electron-positron pairs is the Majorana nature 
%of the neutralino 
%coupled with the fact that the annihilation takes place essentially at rest. 
%Selection rules 
%then force the coupling to fermions to be proportional to the fermion mass. 

There has recently been a new  balloon-borne  detection experiment 
\cite{Barwick}, 
with increased sensitivity to eventual positrons from neutralino annihilation,
where an excess of positrons over that expected from ordinary sources 
has been found. However, since there are many other possibilities to 
create positrons by astrophysical sources, e.g. near the centre of 
the Milky Way, the interpretation is not yet conclusive. 

 Antiprotons could for some supersymmetric parameters constitute a useful 
signal \cite{antiprotons}, but probably the upcoming space 
experiments \cite{ting} will be needed 
to  disentangle a low-energy signal from the smooth 
cosmic-ray induced background.  For kinematical reasons, antiprotons 
created by pair-production in cosmic ray collisions with interstellar 
gas and dust are born with relatively high energy, whereas 
antiprotons from neutralino annihilation populate also the sub-100 
MeV energy band.
A problem that plagues estimates of the signal strength of both positrons and 
antiprotons is, however, the uncertainty of the galactic propagation model 
and solar wind
 modulation. 
 
 Even allowing for large such systematic effects, the 
 measured antiproton flux gives, however, rather stringent limits on the 
 lifetime of hypothetical $R$-parity violating decaying neutralinos 
 \cite{baltz}.

\subsubsection{Methods with distinct experimental signature}

With these problems of positrons and antiprotons, one would expect that 
problems
 of gamma rays and neutrinos are similar, if they only arise from 
secondary decays in the 
annihilation process. For instance, the gamma ray spectrum arising from
the fragmentation of fermion and gauge boson final states is quite 
featureless and gives the bulk of the gammas at low energy where the
cosmic gamma ray background is severe. Also, the density of 
neutralinos in the halo is not large enough to give a measurable flux 
of secondary neutrinos, unless the dark matter halo is very clumpy.
 However, neutrinos can escape from the centre  
of the Sun or Earth, where 
neutralinos may have been gravitationally trapped and therefore their density 
enhanced. Gamma rays may result from loop-induced 
annihilations 
$\chi\chi\to\gamma\gamma$ \cite{gammaline} or $\chi\chi\to Z\gamma$ 
\cite{zgamma}.

The rates of these processes are difficult to estimate because of 
uncertainties in 
the supersymmetric parameters, cross sections and halo density profile. However, 
in contrast to the other proposed detection methods they have 
the virtue of giving  very 
distinct, ``smoking gun'' signals: high-energy neutrinos from the 
centre of the Earth or 
Sun, or monoenergetic photons with $E_\gamma = m_\chi$ or $E_\gamma = m_\chi
(1-m_{Z}^2/4m_{\chi}^2)$ from the halo.

\subsubsection{Gamma ray lines}

 The detection probability of a gamma line signal depends on
the very poorly known density profile of the dark matter halo.

To illustrate this point, let us consider the characteristic 
angular dependence of 
the gamma-ray intensity from neutralino annihilation in the galactic halo. 
Annihilation of neutralinos in an isothermal halo 
with core radius 
$a$ leads to a gamma-ray flux of
\begin{eqnarray}
     {d{\cal F} \over {d \Omega}}\simeq \,
     (2\times10^{-11} {\rm cm}^{-2} {\rm s}^{-1} {\rm sr}^{-1})\times &
     &\nonumber\\ 
    {(\sigma_{\gamma\gamma} v)_{29}
     (\rho_\chi^{0.3})^2\over (m_\chi/\, 10\,
     {\rm GeV})^2} \,{\left(R\over 8.5\ {\rm kpc}\right)}J(\Psi)&&
\end{eqnarray}
where
 $(\sigma_{\gamma\gamma}
v)_{29}$ is the annihilation rate in units of 
     $10^{-29}\, {\rm cm}^3\,{\rm s}^{-1}$,
$\rho_\chi^{0.3}$ is the
local neutralino halo density in units of 0.3 GeV cm$^{-3}$  
and $R$ is the distance 
to the galactic center.
The integral $J(\Psi)$ is given by
\beq
J(\Psi)={1\over R\rho_{0}^2}\int_{\rm 
line-of-sight}\rho^2(\ell)d\ell(\Psi),
\eeq
and is evidently very  sensitive to local density variations along the 
line-of-sight path of integration.

 We remind of the 
fact that since the 
neutralino velocities in the halo are of the order of 10$^{-3}$ of the 
velocity of light, the 
annihilation can be considered to be at rest. The resulting gamma ray 
spectrum is a line 
at $E_\gamma=m_\chi$ of relative linewidth 10$^{-3}$ which in 
favourable cases 
will stand out against background. The process $\chi\chi\to Z\gamma$ 
is treated analogously and has a  similar rate 
\cite{zgamma}.

To compute $J(\Psi)$, a model of the dark matter halo has to be chosen. 
Recently, N-body simulations have given a clue to the final halo 
profile obtained by hierarchical clustering in a CDM scenario 
\cite{NFW}. It turns out that the universal halo profile found in 
these simulations has a rather significant enhancement $\propto 1/r$ 
near the halo centre. If applicable to the Milky Way, this
 would lead to a much enhanced annihilation 
rate towards the galactic centre, and also to a very characteristic 
angular dependence of the line signal. This would be very beneficial 
when discriminating against the galactic and extragalactic $\gamma$ 
ray background, and Air Cherenkov Telescopes (ACTs) would be eminently 
suited to look for these signals, if the energy resolution is at the 
$10-20$ \% level.

The calculation of the $\chi\chi\to\gamma\gamma$ cross section is 
technically quite 
involved with a large number of loop diagrams contributing. In fact, 
only very recently a 
full calculation in the MSSM
was performed \cite{newgamma}.  
Since the 
different contributions all have to be added coherently, there may be 
cancellations or 
enhancements, depending on the supersymmetric parameters.

An important contribution, especially for neutralinos that contain
a fair fraction of Higgsino components, is from $W^+W^-$ intermediate states.
This is also true for the $Z\gamma$ final
state for very massive neutralinos \cite{zgamma}. In fact, thanks to 
the effects of coannihilations \cite{coann}, neutralinos as heavy as 
several TeV are allowed without giving a too large $\Omega$. These 
extremely heavy dark matter candidates (which would require quite a 
degree of finetuning in most susy models) are predominantly higgsinos 
and have a remarkably large branching ratio into the loop-induced 
$\gamma\gamma$ and $Z\gamma$ final states (the sum of these can be as 
large as 30 \%). Recently, there has been some interest in TeV 
neutralinos due to a claim of a possible structure in existing data 
\cite{strausz}. It seems, however, that this claim was based on an 
erroneous estimate of the acceptance of the experiments. Also, the 
purported rate is at least 3 or 4  orders of magnitude larger than 
what can be obtained in susy models \cite{newgamma}.
 
In Fig.\,~1, we show the gamma ray line flux given in a scan
 of supersymmetric
models consistent will all experimental bounds (including $b\to s\gamma$),
assuming an effective value of $10^3$ for the average of $J(\Psi)$ 
over the $10^{-3}$ steradians that typically an Air Cherenkov 
Telescope (ACT) would cover.  (See \cite{BBU} for details.)

\begin{figure}[!htb]
\epsfig{file=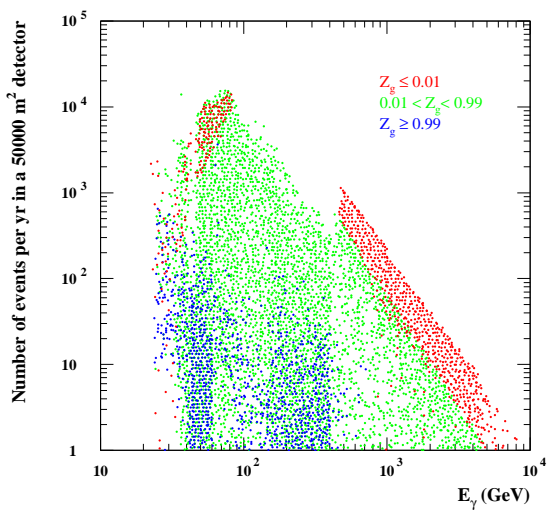,width=8cm}
\caption{Results for the gamma ray line flux in an extensive 
scan of supersymmetric 
parameter space in the MSSM \protect\cite{BBU}. 
Shown is the number of events versus photon energy in an Air Cherenkov Telescope of area $5\cdot 10^4$
m$^2$  viewing the galactic centre for one year. The 
halo profile of \protect\cite{NFW} for the dark matter has been assumed.
The different colours represent different values of the gaugino (photino plus
zino) fraction $Z_g$.}
\label{NC}
\end{figure}

It can be seen that the models which give 
the highest rates should be within reach of the new generation of ACTs 
presently being constructed. These will have an effective area of almost $10^5$
m$^2$, a threshold of some tens of GeV and an energy resolution 
approaching 10 \%. In favourable cases, especially at the low 
$m_{\chi}$ end, also a smaller area detector with  better energy 
resolution and wider angular acceptance such as the proposed GLAST 
satellite could reach discovery potential.

\subsubsection{Indirect detection through neutrinos}

Another promising indirect detection method is to use neutrinos from 
annihilations of neutralinos accumulated in the centre of the Sun or Earth.
This will be a field of extensive experimental investigations in view
of the new neutrino telescopes (AMANDA, Baikal, NESTOR, ANTARES) 
planned or under construction \cite{halzen}.

The capture rate induced by scalar (spin-independent) interactions between
the neutralinos and the nuclei in the interior of the Earth or Sun is 
the most difficult one to compute, since it depends sensitively on
Higgs mass, form factors, and other poorly known quantities. For the Sun,
the axial cross section is relatively easy to compute, a good approximation
is given by \cite{jkg}
\beq
     C_{\rm ax}^\odot = (1.3\times10^{25}\, {\rm s}^{-1})\,
         {\rho_{0.3}^\chi \,
     \sigma_{0\,spin}^{H(40)} 
     \over (m_\chi/(1 {\rm GeV}))  \bar v_{270}}
\eeq
where $\sigma_{0\,spin}^{H(40)}$ is the cross section for
neutralino-proton elastic scattering via the axial-vector interaction
in units of $10^{-40} {\rm cm}^2$, $\bar v_{270}$ is the dark-matter
velocity dispersion in units of 270 ${\rm km} {\rm s}^{-1}$, and
$\rho^\chi_{0.3}$ is the local halo mass density in units of 
$0.3$ GeV cm$^{-3}$. The capture rate in the Earth is dominated
by scalar interactions, where there may be kinematic and other
enhancements, in particular if the mass of the neutralino almost
matches one of the heavy elements in the Earth. For this case,
a more detailed analysis is called for, but convenient approximations
are available \cite{jkg}.

To illustrate the potential of neutrino telescopes for discovery of dark matter
through neutrinos from the Earth or Sun, we present the results of a
full calculation \cite{BEG}. In Fig.\,~2 it can be seen that a neutrino
telescope of area around 1 km$^2$, which is a size currently being discussed,
would have discovery potential for a range of  supersymmetric models.

\begin{figure}[!h]
         \epsfig{file=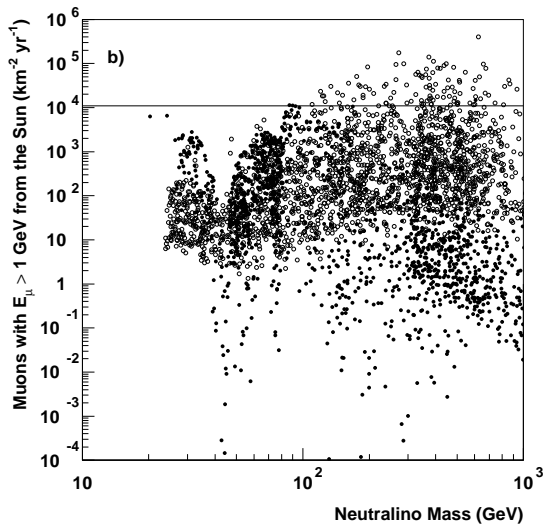,width=8cm}
         \caption{The indirect detection rates from
  neutralino annihilations in the Sun  versus the
  neutralino mass. The
  horizontal line is the Baksan limit \protect\cite{baksan}. For 
  details, see \protect\cite{BEG}.}
         \label{sun}
\end{figure}

If a signal were established, one can use the angular spread caused by
the radial distribution of neutralinos (in the Earth) and by the 
energy-dependent mismatch between the direction of the muon and that 
of the neutrino (for both the Sun and the Earth) to get a rather good 
estimate of the neutralino mass \cite{EG}. If muon energy can also be 
measured, one can do even better \cite{BEK}.

\section{Conclusions}

To conclude,  indirect detection methods have the potential
to be very useful complements to direct detection of supersymmetric dark
matter candidates. In particular, new air Cherenkov and neutrino telescopes
may have the sensitivity to rule out or confirm the supersymmetry 
solution of the dark matter problem. 

Since also the ex\-perimental situa\-tion con\-cerning massive neu\-trinos and 
axions is getting clearer, there is a chance to reach the goal of 
explaining the nature of the dark matter in the not too distant 
future.
\section{Acknowledgements}
I wish to thank my collaborators, in particular Joakim Edsj\"o, Paolo Gondolo 
and
Piero Ullio, for many
helpful discussions. This work has been supported in part by the
Swedish Natural Science Research Council (NFR) and the European Union
Theoretical Astroparticle Physics Network (TAN).

\end{document}